\documentclass{iscram}%
\usepackage[utf8]{inputenc}

\iscramset{
title={%
    Analysing Donors' Behaviour in \\Non-profit Organisations for \\Disaster Resilience \\{\LARGE The 2019--2020 Australian Bushfires Case Study}%
},
short title={%
    Analysing Donors' Behaviour in Non-profit Organisations for Disaster Resilience%
},
author={%
    short name={D.\ Rajapaksha},
    full name={Dilini Rajapaksha}\thanks{Equal contribution.},
    affiliation={ARC Centre of Excellence for Automated Decision-Making and Society, \\RMIT University, Australia}%
},
author={%
    short name={K.\ Sokol},
    full name={Kacper Sokol}\footnotemark[1],%
    affiliation={ARC Centre of Excellence for Automated Decision-Making and Society, \\RMIT University, Australia}%
},
author={%
    short name={J.\ Chan} ,
    full name={Jeffrey Chan},
    affiliation={ARC Centre of Excellence for Automated Decision-Making and Society, \\RMIT University, Australia}%
},
author={%
    short name={F.\ Salim}, 
    full name={Flora Salim},
    affiliation={ARC Centre of Excellence for Automated Decision-Making and Society, \\UNSW  Sydney, Australia}%
},
author={
    short name={M.\ Prasad}, 
    full name={Mukesh Prasad},
    affiliation={School of Computer Science, FEIT, \\University of Technology Sydney, Australia}%
},
author={
    short name={M.\ Samarawickrama},
    full name={Mahendra Samarawickrama},
    affiliation={Australian Red Cross}%
},%
footer/line 1={%
    CoRe Research \emph{Paper} -- Social Media for Disaster Response%
},%
footer/line 2={%
    \emph{Proceedings of the ISCRAM Asia Pacific Conference 2022}%
},%
footer/line 3={%
    Thomas Huggins and Vincent Lemiale eds%
}%
}

\usepackage{enumitem}
\usepackage{wrapfig}
\usepackage{placeins}
\usepackage{todonotes}
\usepackage{booktabs}
\usepackage{url}
\usepackage{svg}
\usepackage{subcaption}
\usepackage{graphicx}
\usepackage{hyperref} %
\let\oldcite\cite
\newcommand{\citet}[1]{\oldcite{#1}}
\renewcommand{\cite}[1]{\parencite{#1}}

\begin{document}

\maketitle

\abstract{%
With the advancement and proliferation of technology, non-profit organisations have embraced social media platforms to improve their operational capabilities through brand advocacy, among many other strategies. The effect of such social media campaigns on these institutions, however, remains largely underexplored, especially during disaster periods. This work introduces and applies a quantitative investigative framework to understand how social media influence the behaviour of donors and their usage of these platforms throughout (natural) disasters. More specifically, we explore how on-line engagement -- as captured by Facebook interactions and Google search trends -- corresponds to the donors' behaviour during the catastrophic 2019--2020 Australian bushfire season. To discover this relationship, we analyse the record of donations made to the Australian Red Cross throughout this period. Our exploratory study reveals that social media campaigns are effective in encouraging on-line donations made via a dedicated website. We also compare this mode of giving to more regular, direct deposit gifting.%
}


\section{Introduction}
Nowadays, social media are widely adopted to improve disaster readiness and resilience. %
Specifically, the utilisation and usage patterns of such platforms have been studied for different aspects of emergency management; for example, organisational decision-making, disaster management, strategic planning for emergency response, as well as the overall influence of social media on relevant organisations and their activities. Among these avenues of inquiry, Non-Profit Organisations (NPOs) are often keen to learn how to acquire new donors and mobilise existing givers to increase the organisational capacity, which is urgently needed during disaster periods. For such NPOs, the donations collected from \emph{individuals} are usually the lifeline necessary to ``survive, sustain themselves, and develop''~\cite{medina2014modeling}. While the number of fundraising agencies has soared in this fast-paced environment, the quantity of donors and donations reaching any single organisation have decreased as a consequence~\cite{ford2010nostalgia}. This is especially true for individuals who -- among corporations, foundations and governments -- generate a sizeable portion of NPOs' budget through their monetary contributions.%

Individual, goodwill donations are a precarious source of relief money managed by NPOs. %
It is therefore worth better understanding various factors influencing such non-institutional giving.
\citet{nah2009media} suggests that the Internet is the most appropriate tool for reaching out to the broader population of potential donors in the connected world. %
Specifically, this technology allows organisations to publish a website and, more recently, venture into commercial social networking sites such as Facebook and Twitter by creating a page on the former and curating a feed on the latter. %
NPOs should devise strategies to lead people to those social media accounts and encourage them to actively participate in these spaces.
Using the example of Facebook, such interactions are mostly captured by \emph{likes}, \emph{shares} or \emph{comments} attached to posts published thereon by a stakeholder.
Engaging with individual donors to encourage active participation across the NPOs' social media portals has the potential to boost the overall outreach of these organisations -- via voluntary brand advocacy~\cite{wilk2018navigating}, among others -- which can also lead to increased participation in off-line activities such as in-person donating, hosting fundraisers and volunteering~\cite{levenshus2010online}.%

In this study, we aim to analyse the effects of various activities happening across social media on non-profit organisations during natural disasters. %
To this end, we investigate the donation data collected by the Australian Red Cross during the 2019--2020 bushfire season.\footnote{~The data cannot be made available to the public due to their sensitivity.}
Throughout this calamity, the charity supported 49,718 people affected by the fires, awarded 5,914 people with bushfire grants (AU\$187m), and extended the recovery programme to 21,563 people, which amounted to AU\$207m throughout that period~\cite{redcross}. %
Given the overall success of these actions, it is crucial to explore the effects of social media activities on the collected donations to better prepare for the future.

\section{Related Work}%

Social media can be used to various ends in case of disasters; for example, communication, fundraising and publicising relief efforts~\cite{finau2018social}. %
To formalise these roles, \citet{houston2015social} offered a functional framework for social media use across various stages and aspects of disasters.
Among others, the authors identified the important role of such platforms in enabling their users to \emph{donate}, and charitable organisations to \emph{receive donations}, both during and after the event of interest. %
Specifically for Facebook, \citet{bhati2020success} showed that fundraising success -- determined by the number of donors and the value of their donations -- is linked to the non-profit's number of likes (network size), posts (activity) and shares (audience engagement).
Additionally, \citet{wilk2018navigating} reported that customers -- i.e., supporters of a charity who donate -- have more impact on consumer (i.e., donor) decision-making than brand-sponsored marketing messages thanks to a process called \emph{voluntary brand advocacy}. %
More specifically, people who mention a charity on social media and engage in brand advocacy (on Facebook) are more likely to donate~\cite{wallace2017does,choi2021developing}.

Similarly, \citet{mathur2019building} showed that social media brand advocacy and reciprocity are crucial for improving brand equity, formalising this process in a conceptual framework. %
\citet{okada2017effectiveness}, on the other hand, scrutinised the effectiveness of social media in encouraging people to donate in view of disasters.
In particular, their findings demonstrated that employing social media both \emph{before} and \emph{after} a disaster increases the amount of collected donations.
Additionally, the authors found that the use of social media allows to raise more money in comparison to in-house platforms such as websites and blogs.
\citet{okada2017effectiveness} have also speculated that the \emph{first three months} of a disaster may be the most effective period to collect donations.

\section{Background}

Before we delve into our results, we define relevant terms and overview key concepts.%
\begin{description}[leftmargin=20pt,itemindent=0pt,labelsep=8pt,align=left,font={\it}]
\item [Facebook] Free-to-use, on-line social networking platform started in 2004 for college students~\cite{lee2012young}.
\item [Google Analytics] A toolkit that allows website publishers to determine the number of visitors to their sites. Additionally, it can measure the performance of marketing channels, sources of traffic and success of (advertising) campaigns over time.%
\item [Non-profit organisations] Groups established for purposes other than statutory or profit-maximising, and in which no part of the organisation's income is distributed among its members, directors or officers~\cite{maier2016nonprofit}. NPOs are also known as \emph{non-profit corporations}, \emph{not-for-profit agencies} or \emph{non-profit institutions}.%
\item [Charity organisations] A subgroup of NPOs that strives to improve various aspects of life across different communities. These organisations follow strict guidelines for qualification and must be fully registered to be classified as a charity. Achieving this status grants them an authority to issue official receipts for income tax deductions, which allows them to receive tax credits~\cite{dimaggio1990sociology}.%
\item [Social media] Any on-line resource that is designed to facilitate engagement between individuals~\cite{aichner2021twenty}.
\end{description}

\paragraph{The role of social media in NPOs}%
Non-profit organisations can benefit from social media in different ways, such as attracting new volunteers or donors, advertising events or actions to the broader community, and building new relationships. \citet{levine2012online} have discovered a positive link between on-line media presence and higher market orientation, which may boost financial viability. This dependency could, for example, be observed when Red Cross raised \$32 million in donations for Haiti using social media alone~\cite{daniels2010nonprofits}. Moreover, \citet{feng2017social} reported that social media platforms help to enhance trust and satisfaction in NPOs. This was validated in a study carried out by \citet{goldkind2015social}, who found that NPOs using social media were able to increase the interest in their organisations up to 400\% over a year, ultimately helping them to boost their brand advocacy.%

\paragraph{Natural disasters}%
Natural disasters are sudden catastrophic phenomena that can occur all over the world. The severity level of such an event depends on the strength of the disaster, probability of its recurrence, location, population density, general economic development of the affected area, preparedness of the community as well as its response to the disaster. The destruction caused by such events is often far-reaching, costly to mitigate, and it creates a huge demand for the essential goods and services, all of which require a rapid response. In particular, to effectively overcome such a situation it is usually necessary to seek aid from both %
governments and the non-profit sector. \citet{mckenzie2011effects} further corroborated this observation by uncovering that donations from the members of the public tend to be greater than government contributions following the occurrence of a natural disaster. Moreover, the author concluded that the media coverage associated with these events usually generates the most significant increase in donations flowing into NPOs. It is therefore worthwhile exploring how on-line platforms like social media portals and search engines are linked to the behaviour of donors.%
 
\begin{figure}[t]
    \centering
    \includegraphics[width=0.85\textwidth]{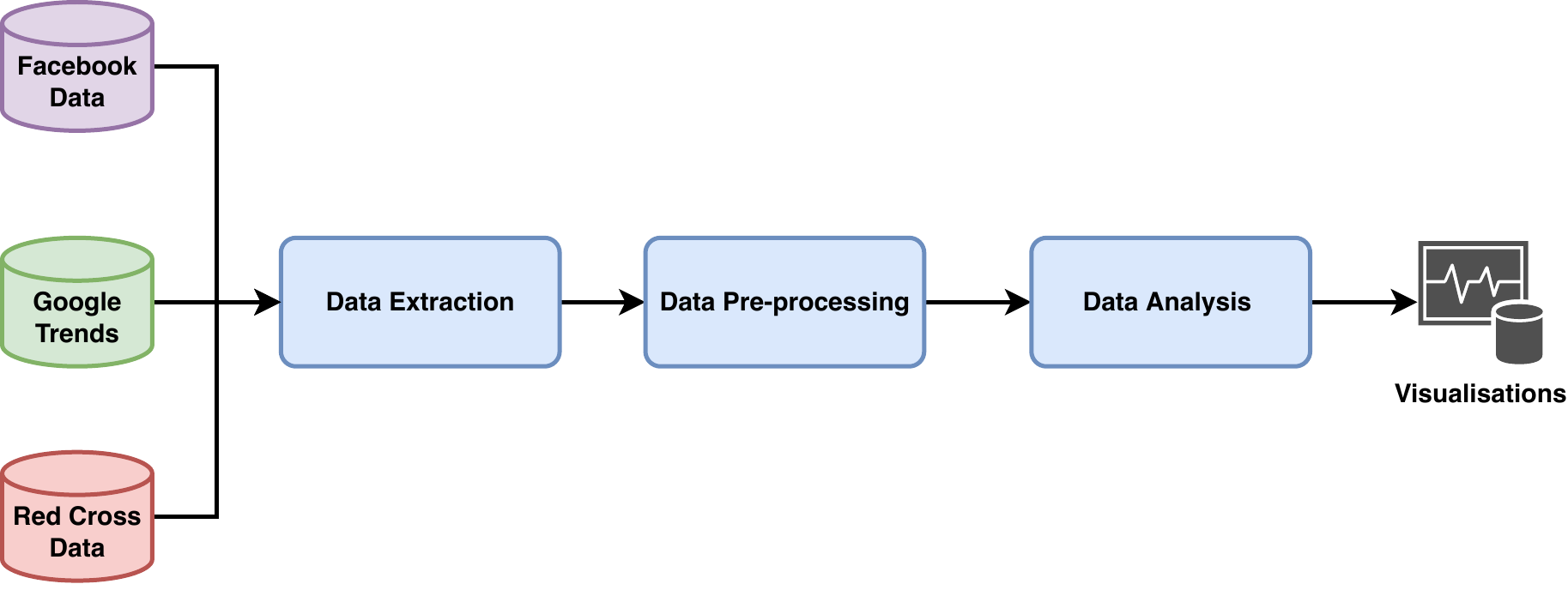}%
    \caption{High-level overview of the analysis applied to the 2019--2020 Australian bushfire disaster data.}
    \label{Fig:high_level_diagram}
\end{figure}

\section{Methodology and Data Sets}

In this section, first, we discuss the workflow employed to extract the data used for our study. Next, we explain the pre-processing steps applied to these data sets.
We then, in the following section, aggregate the resulting data to investigate multiple scenarios throughout the disastrous 2019--2020 Australian bushfire season. %
An overview of this process is depicted in Figure~\ref{Fig:high_level_diagram}.%

In this study we consider three sources of data related to the 2019--2020 bushfire season in Australia; namely, Facebook activity data, Google Trends and donation data provided by the Australian Red Cross. Since our investigation aims to analyse the influence of social media and other on-line presence on NPOs during the aforementioned bushfire season, we consider a period of time when the donation collection coincided with relevant media campaigns. Before presenting the results of our analysis, we discuss the data extraction and pre-processing steps applied to each data source.%

\paragraph{Facebook Data}
To gather insights from Facebook, we focus on various social media engagement metrics pertaining to the Australian Red Cross' Facebook page\footnote{~\url{https://www.facebook.com/AustralianRedCross}} in the period between October 2019 and March 2020. To extract these data, we use an open source tool called Facepager\footnote{~\url{https://github.com/strohne/Facepager}}. We then apply \texttt{QUERY = bushfire OR fire} to this time series to extract posts related to bushfires. %
Before analysing the data, we remove all the \texttt{null} and \texttt{NA} values. Next, we aggregate the data based on the daily count of reactions, shares and likes. After these steps, we are left with 63 records, which we use to analyse the behaviour of users on social media during the 2019--2020 Australian bushfire season.%

\begin{figure}[t]
    \centering
    \includegraphics[width=0.75\textwidth]{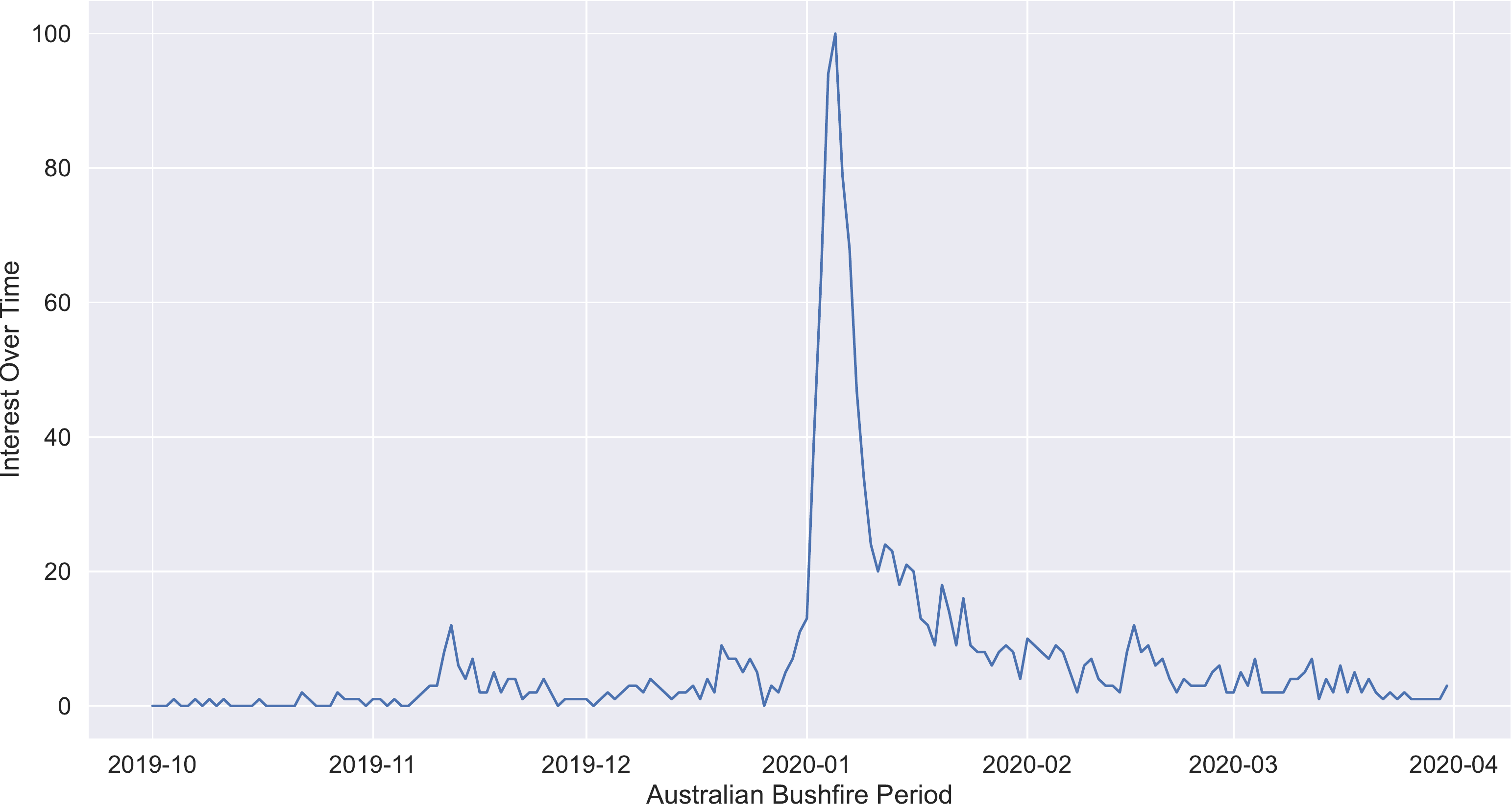}
    \caption{Google search trend for the \texttt{bushfire} keyword in Australia during the disaster period. The y-axis represents search interest relative to the highest point visible in the graph for Australia between the 1\textsuperscript{st} of October 2019 and the 31\textsuperscript{st} of March 2020 (shown on the x-axis). A value of 100 signifies the peak popularity of the term. A score of 50 denotes that the keyword is half as popular. A value of 0 indicates lack of (enough) data for this particular term.}%
    \label{Fig:bushfire_google_trend}
\end{figure}

\paragraph{Google Trends}%
We consider Google Trends to understand the on-line awareness of people about the Australian bushfires during the period between October 2019 and March 2020. To perform the analysis we select \texttt{Australia} as the country and consider \texttt{QUERY = bushfire} as the filtering criterion. Figure~\ref{Fig:bushfire_google_trend} shows the search trend for the \texttt{bushfire} keyword during the disaster period.%

\begin{wraptable}{r}{7cm}%
    \centering
    \begin{tabular}{rr}
    \toprule%
     {\bf Donation Method} & {\bf Number of Donations} \\ %
    \midrule%
     direct deposit & 3,086  \\  
     mail & 1,223 \\
     telephone  & 60\\
     website & 699,562 \\
    \bottomrule%
    \end{tabular}
    \caption{Number of donations made via each available route during 2019--2020 Australian bushfires.\label{Tab:response_donation}}%
\end{wraptable}
\paragraph{Red Cross Data}
We consider a time series capturing donation patterns across the bushfire fundraising campaigns organised by the Australian Red Cross between 2019 and 2020. %
The raw data set consists of instances relating to multiple donation drives. In this analysis, first, we extract the data between October 2019 and March 2020. Next, we use \texttt{QUERY = bushfire OR evacuate OR survivors OR fire} to narrow down the scope of the records to the ones pertaining to bushfires. %
Afterwards, we pre-process the data by removing the \texttt{NULL} and \texttt{NA} values. We then aggregate the donations into a 24-hour resolution to visualise the daily donation rate and the amount of money collected across the bushfire period. After these steps 703,931 donation records remain and are considered in our analysis. The breakdown of the donations based on the gifting method is shown in Table~\ref{Tab:response_donation}.%

\section{Data Analysis and Insights}%

Our study is focused primarily on investigating the number of on-line and off-line donations in relation to various events and activities recorded across Facebook and Google Trends throughout the bushfire period. %
Notably, the Australian Red Cross considers this particular fundraising drive as a \emph{single giving-tied programme}, which means that \emph{donation pledges} are not accounted for in the analysed data set. %
Therefore, for \emph{off-line} contributions our study only considers donations made via \emph{direct deposit}, \emph{mail} or \emph{telephone}. %
Additionally, we analyse \emph{on-line} gifting facilitated through the Australian Red Cross' website.%

\begin{figure}[t]%
    \centering
    \begin{subfigure}[b]{\textwidth}
    \centering
    \includegraphics[width=0.75\textwidth]{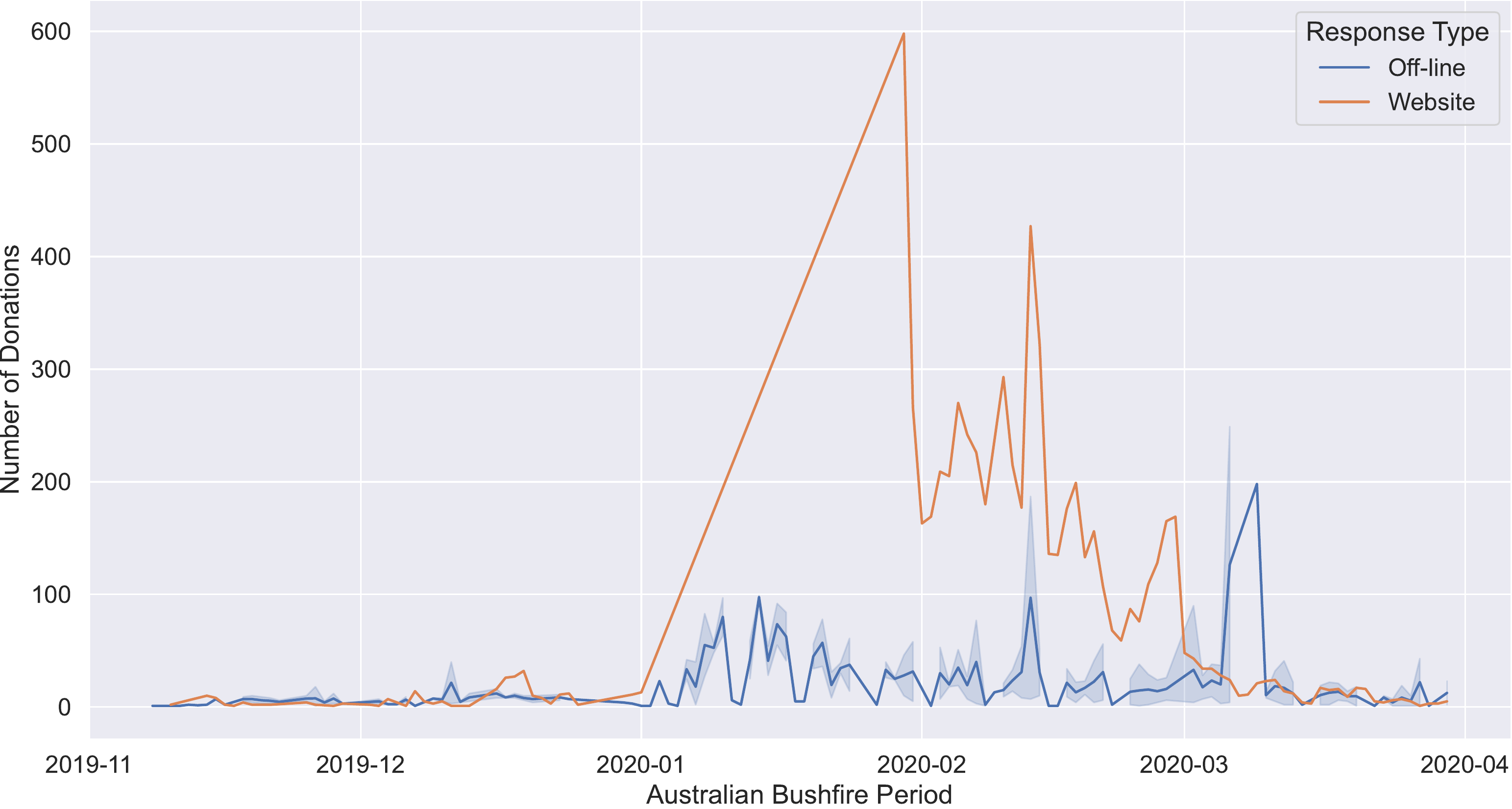}
    \caption{Number of on-line and off-line donations during the selected bushfire period. The orange line indicates the count of donations made via the website, whereas the blue line denotes the number of donations made through off-line means (direct deposit, mail and telephone). We only consider days with fewer than 1,500 donations to better capture the variability of the off-line donations.}%
    \label{Fig:online_offline_count}
    \end{subfigure}\\[1em]
    \begin{subfigure}[b]{\textwidth}
    \centering
    \includegraphics[width=0.75\textwidth]{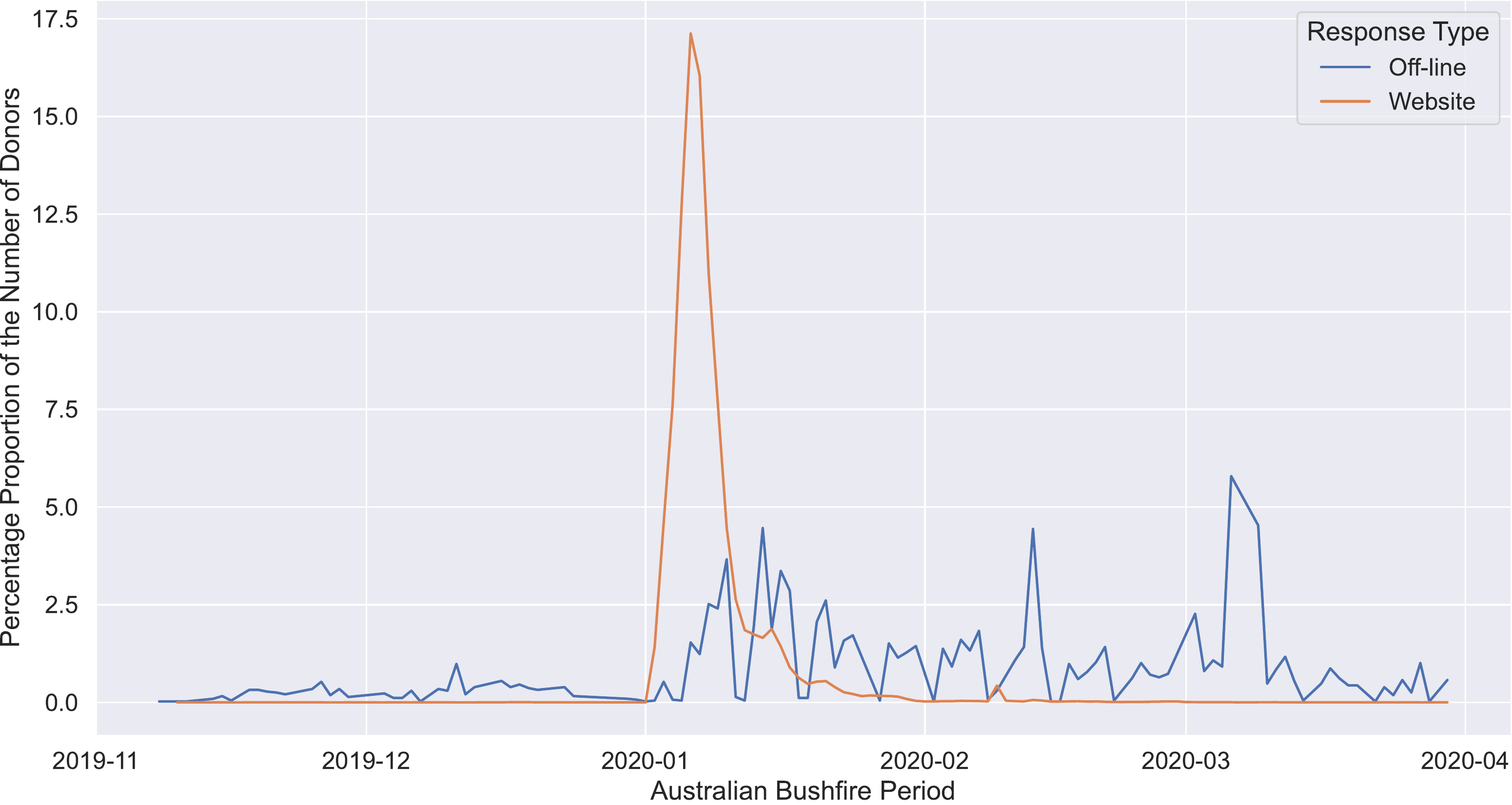}
    \caption{Percentage proportion of on-line and off-line donors during the selected bushfire period. The orange line charts the proportion of donors using the website, whereas the blue line shows the proportion of donors using off-line means (direct deposit, mail and telephone).}%
    \label{Fig:website_offline_count_pct}
    \end{subfigure}
    \caption{Visualisation of the (\subref{Fig:online_offline_count}) distribution of donations and the (\subref{Fig:website_offline_count_pct}) behaviour of donors.}
    \label{Fig:donations}
\end{figure}

\begin{figure}%
    \centering
    \includegraphics[width=0.75\textwidth]{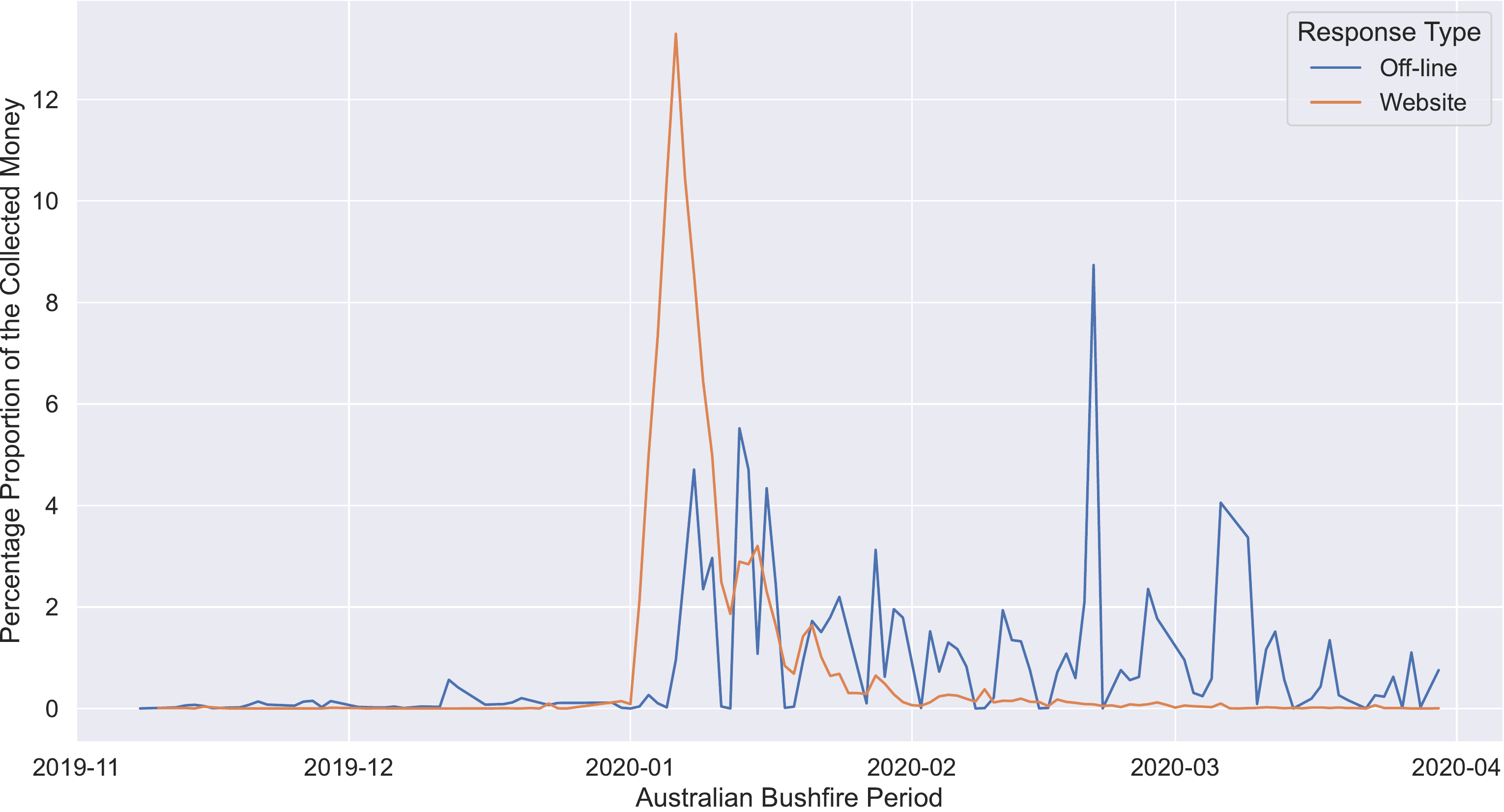}
    \caption{%
Percentage proportion of the on-line and off-line donation amount across the selected bushfire period.
The orange line indicates the proportion of the total collected amount across the displayed time for website donations, whereas the blue line captures the same metric for off-line transactions (direct deposit, mail and telephone).}%
    \label{Fig:online_offline_sum}
\end{figure}

The behaviour of donors and influx of donations are summarised in Figure~\ref{Fig:donations}. %
More specifically, Panel~(\subref{Fig:online_offline_count}) visualises the \emph{number of donations} made throughout the period of interest; note that the plot only displays the days with fewer than 1,500 donations to ensure the readability of the figure, thus better convey the variability of off-line contributions. %
Panel~(\subref{Fig:website_offline_count_pct}), on the other hand, illustrates the percentage proportion of the number of donors over the same time period. %
Figure~\ref{Fig:donations} clearly shows that most of the donations were received via the Australian Red Cross' website, especially so between January and February. %
Additionally, the distribution of the influx of money over the period of interest -- shown in Figure~\ref{Fig:online_offline_sum} as the percentage proportion of the total amount of donations collected in that time span to avoid revealing sensitive information -- uncovers that the value of contributions is higher for \emph{website} donations only throughout January (note that outliers were removed to improve the readability of this chart). %
Based on the same plot, the Australian Red Cross was able to attract more funding via \emph{off-line} donation routes in the remaining period.%

\begin{figure}%
    \centering
    \subfloat[
    Relation between Facebook interactions and Google searches, and the percentage proportion of the number of donors contributing either via direct deposit or through the Australian Red Cross' website during the selected bushfire period. %
    The bar chart shows the percentage proportion of on-line engagement captured by Google Trends as well as Facebook shares and reactions (the legend and y-axis to the right). %
    The green and red lines indicate the percentage proportion of the number of donors contributing respectively through direct deposit and website (the legend and y-axis to the left).%
    ]{
    \includegraphics[width=0.75\textwidth]{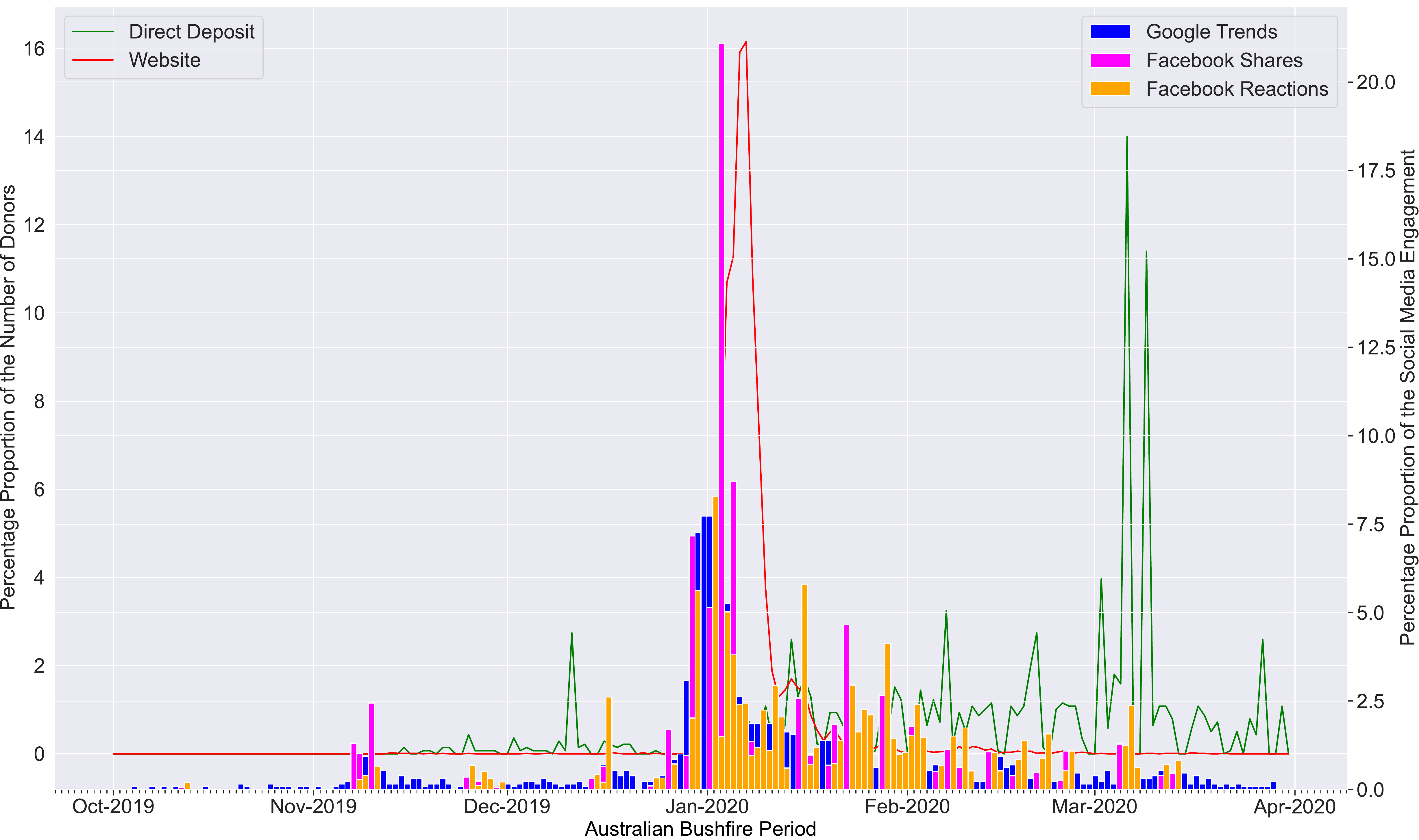}\label{Fig:online_offline_detailed_count_pct_joint}}\\[1em]
    \subfloat[
    Decomposition of Panel~(\subref{Fig:online_offline_detailed_count_pct_joint}) into two individual plots for improved readability.
    ]{
    \includegraphics[width=0.485\textwidth]{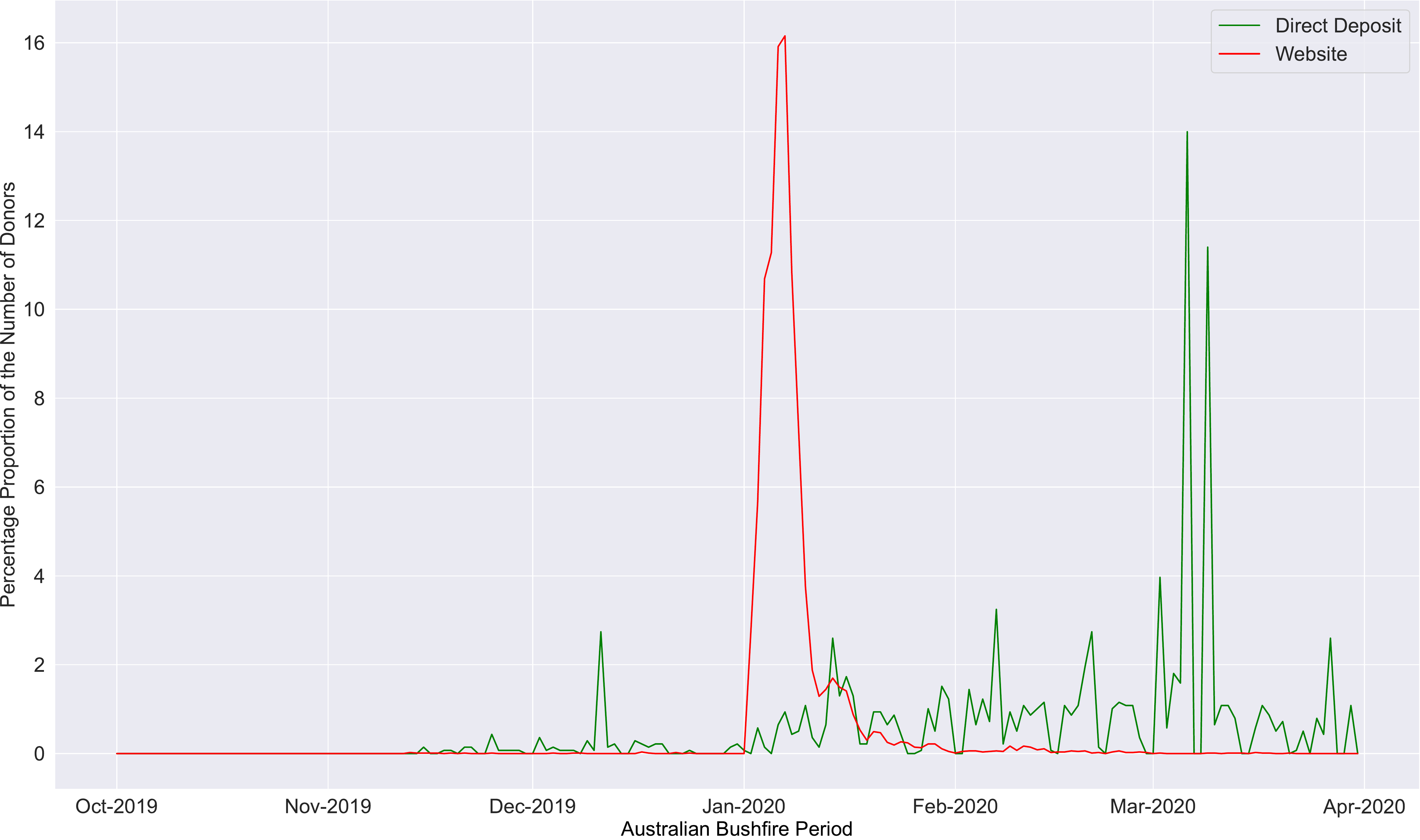} \hfill
    \includegraphics[width=0.485\textwidth]{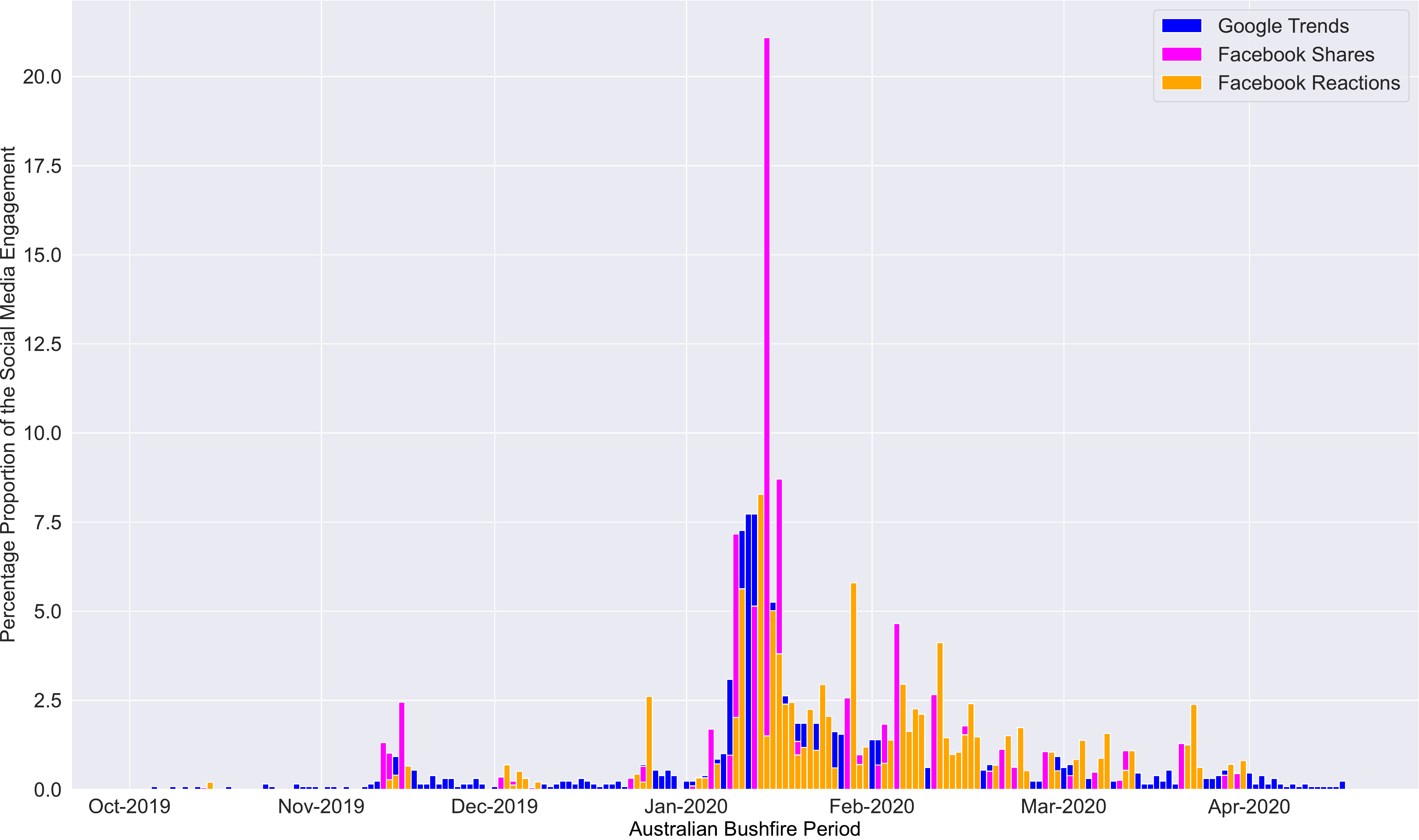}
    \label{Fig:online_offline_no_of_donations}}
    \caption{Relation between Facebook interactions and Google searches, and the number of donors contributing either via direct deposit or through the Australian Red Cross' website during the selected bushfire period.}
    \label{Fig:online_offline_detailed_count_pct}
\end{figure}

Next, we shift our attention to the sudden spike in the number (Figure~\ref{Fig:online_offline_count}) and value (Figure~\ref{Fig:online_offline_sum}) of monetary contributions made through the website across January. %
According to the schedule of public awareness and fundraising campaigns as well as the calendar of disaster relief and recovery efforts shared with us by the Australian Red Cross, their significant portion was executed between the 31\textsuperscript{st} of December 2019 and the 31\textsuperscript{st} of January 2020. %
This corresponds to, and explains, the aforementioned rise in donation metrics. %
Notably, smaller social media campaigns pertaining to the bushfires were run up to the 17\textsuperscript{th} of February 2020; we hypothesise that they may be the reason for a higher number of on-line (website) donations when compared to off-line giving until the end of February (Figure~\ref{Fig:online_offline_count}).%

\begin{figure}[t]
    \centering
    \subfloat[
    Relation between Facebook interactions and Google searches, and the percentage proportion of the amount of money collected either via direct deposit or through the Australian Red Cross' website during the selected bushfire period.
    The bar chart shows the percentage proportion of on-line engagement captured by Google Trends as well as Facebook shares and reactions (the legend and y-axis to the right).
    The green and red lines indicate the percentage proportion of the amount of money collected respectively through direct deposit and website (the legend and y-axis to the left).
    ]{
    \includegraphics[width=0.75\textwidth]{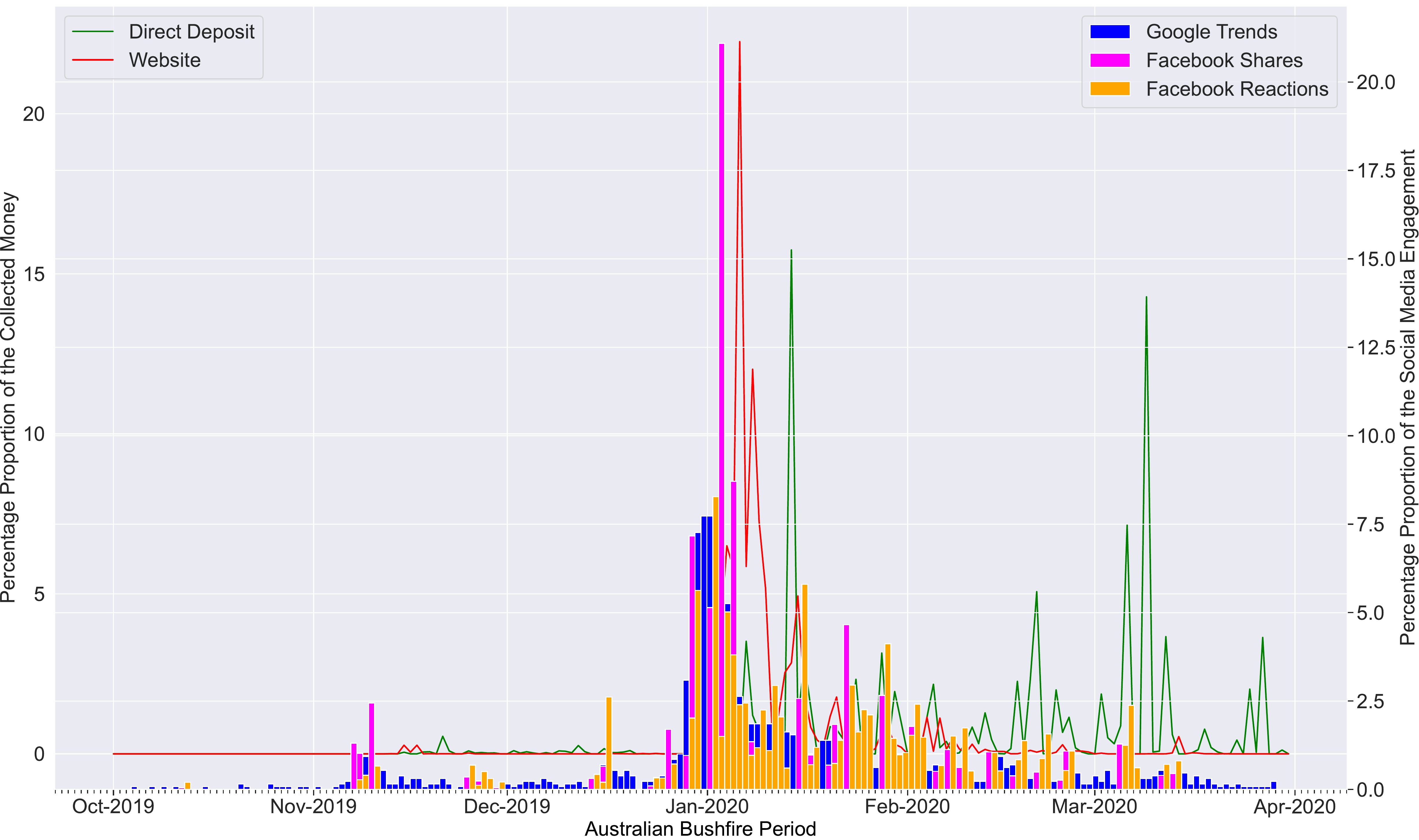}
\label{Fig:online_offline_detailed_sum_pct_joint}}\\[1em]
    \subfloat[
    Decomposition of Panel~(\subref{Fig:online_offline_detailed_sum_pct_joint}) into two individual plots for improved readability.
    ]{
    \includegraphics[width=0.485\textwidth]{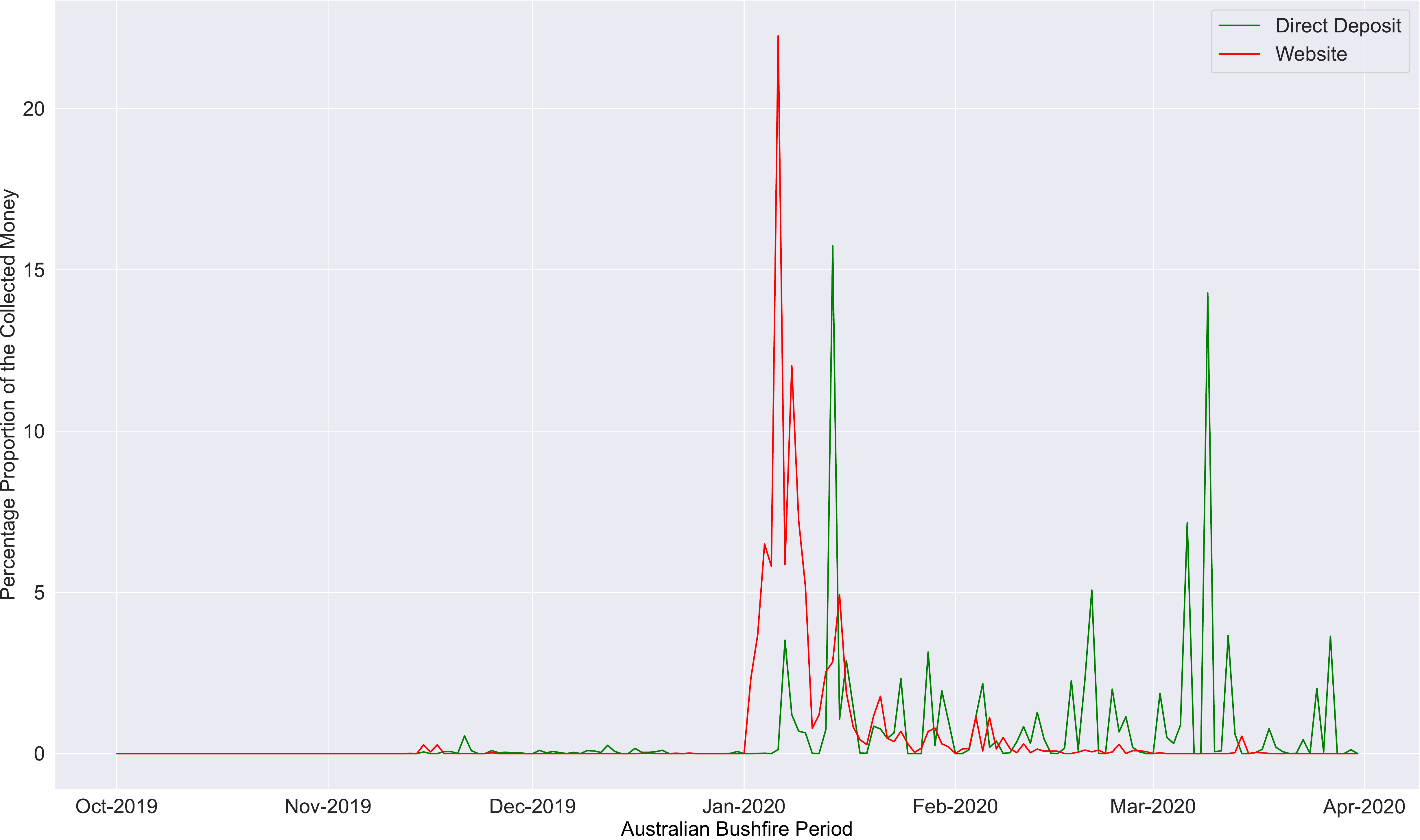} \hfill
    \includegraphics[width=0.485\textwidth]{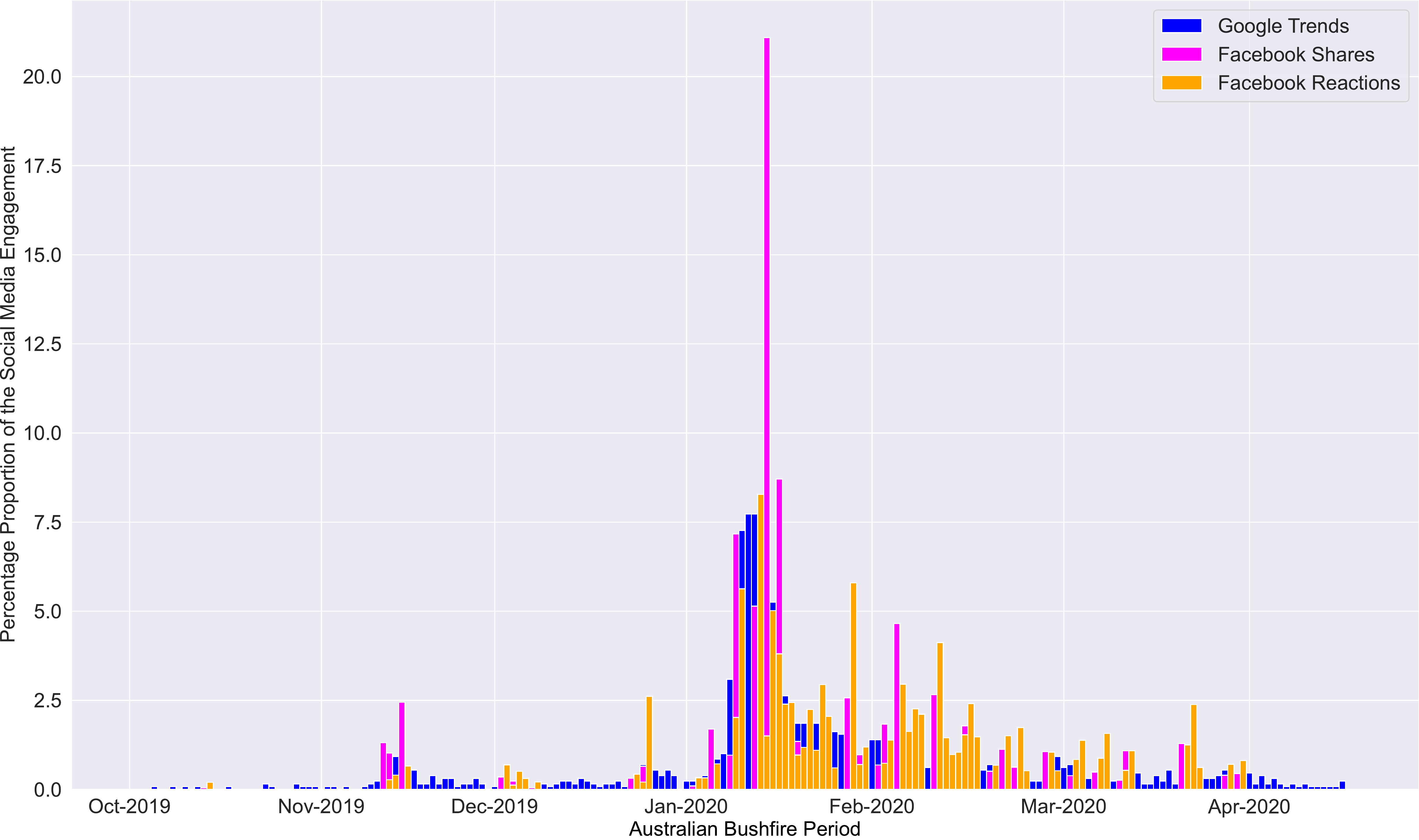}
    \label{Fig:online_offline_sum_of_donations}}
    \caption{
    Relation between Facebook interactions and Google searches, and the amount of money collected either via direct deposit or through the Australian Red Cross' website during the selected bushfire period.
    }
    \label{Fig:online_offline_detailed_sum_pct}
\end{figure}

Another interesting avenue of inquiry is the behaviour of individuals who donate multiple times towards the same cause, specifically bushfire relief efforts coordinated by the Australian Red Cross. %
In this context, we aim to study how on-line (communication and interaction) media may impact such donors, who contribute via recurring direct deposit. %
To this end, we examine the pattern of donations made after the first direct deposit payment, all of which originate from a single person. %
Moreover, we look into the possible influence of on-line media on multiple-time donors who made their first contribution through the website; %
we consider only the first website-based donation to explore this perspective. %
In the Red Cross data set, there are 8,637 individuals who donated multiple times, 8,301 of whom used the website, 333 relied on direct deposit and 3 utilised both options. %
Figures~\ref{Fig:online_offline_detailed_count_pct} and \ref{Fig:online_offline_detailed_sum_pct} illustrate the history of Facebook interactions and Google searches in relation to the number of donors and the amount of money collected throughout the period of interest either via the website or direct deposit.%

A dominating phenomenon in these plots is the aforementioned soaring number of Facebook \emph{shares} and \emph{reaction} attached to entries published on a page curated on this social media platform by the Australian Red Cross, as well as Internet \emph{searches} captured by Google Trends throughout January 2020. %
As explained earlier, these events -- and the co-occurring steep increase in the number and value of on-line (website) donations -- are likely a consequence of the coordinated social media campaigns executed between the 31\textsuperscript{st} of December 2019 and the 31\textsuperscript{st} of January 2020. %
This rise in the engagement of Internet users, especially in terms of Facebook shares, can be interpreted as growth in the NPO's brand advocacy. %
Interestingly, direct deposit donations peak around March 2020, lagging behind more responsive website donors as shown in Figure~\ref{Fig:online_offline_detailed_count_pct}. %
This shift illustrates the \emph{delayed effect} of Facebook marketing campaigns on generation of more sustainable, and possibly regular, off-line donations. %
The sudden increase in the value of off-line donations immediately following the series of social media campaigns visible in Figure~\ref{Fig:online_offline_detailed_sum_pct}, on the other hand, suggests that these communication efforts were likely to nudge some regular donors to additionally contribute with a one-off donation.%

In summary, social media campaigns appear to be a viable and effective route to mobilise donors and encourage them to contribute for the first time or, in case of active benefactors, offer additional monetary support, as well as create new and convert on-off donors to regular givers, all of whom are particularly important to NPOs responding to (natural) disasters. %
Based on our findings, goal-driven Internet presence allows non-profit organisations to boost their brand advocacy and revenue within a short window of time, especially so via on-line donations collected through the NPOs' websites (in contrast to more traditional direct deposit gifting that may be slow to build up). %
More broadly, any social medium and communication platform -- whether on-line or off-line -- seems to benefit the (financial) capability of a non-profit organisation to respond to a calamity by informing the population how to help with or contribute to relief efforts.%

\section{Conclusions and Future Work}

In this study, we explored the relationship between social media engagement as well as other on-line interactions, and the behaviour of donors who contribute money to non-profit organisations during natural disasters. %
Our analysis confirmed that NPOs can benefit from employing social media for various fundraising activities and campaigns; %
additionally, the use of such platforms can stimulate public awareness while also increasing brand advocacy. %
Our investigation considered the disastrous 2019--2020 Australian bushfires, for which we analysed the behaviour of people donating via different routes (on-line, direct deposit, mail and telephone) to the Australian Red Cross. %
Specifically, we compared the patterns of their contributions to user activity on the Australian Red Cross' Facebook page and Google search trends throughout this calamity.%

Our study found that the higher user engagement recorded across social media and other Internet platforms is linked to a greater number of donations made on-line via means such as NPOs' websites. %
More precisely, the social media campaigns executed by the Australian Red Cross during the monitored disaster period helped it to boost the users' on-line engagement, which in turn increased the number of donations made through its website. %
This case study is a prime example of social media's positive influence on NPOs' brand advocacy. %
For off-line donations, on the other hand, it is worth observing a lag between publishing social media campaigns and the donors' response, which is captured by the delayed influx of direct deposit contributions. %
While such donations arrive late, they are still essential for relief efforts during and after a disaster period, especially that they may become regular, hence steady, source of revenue for NPOs. %
Even though on-line donations (made via a website) are more immediate, our analysis revealed that they tend to be unreliable (in contrast to off-line contributions). %
We hypothesise that this pattern arises due to the website donations being mainly driven by social media campaigns and the ensuing engagement of on-line users, which possibly relies on the spontaneous enthusiasm spurred by the content published by NPOs. %
Importantly, social media campaigns help to increase both the number and value of donations received by non-profit organisations throughout disaster periods regardless of the transaction origin.%

In this study we focused exclusively on the relation between on-line platforms, spanning social media as well as other interaction channels available on the Internet, and two effectiveness criteria crucial to any non-profit organisation responding to a (natural) disaster: the behaviour of its donors and viability of its brand advocacy. %
However, it is important to note that traditional media -- such as television and radio broadcasts as well as newspapers -- can influence the aforementioned benchmarks as well. %
In future work, therefore, we plan to expand our investigation to relevant off-line channels, thus offering a more accurate picture of how each communication or interaction medium affects the performance of NPOs during disaster periods. %
Additionally, we envisage validating the insights uncovered here for natural disasters other than bushfires, as well as increasing the granularity of our analysis by separating such periods into three stages -- pre-disaster, disaster and post-disaster -- thus allowing us to understand changes in the behaviour of donors and the engagement of social media users in finer detail. %
All of these findings should inform non-profit organisations how to better manage, utilise and optimise both their on-line and off-line presence, therefore increasing the number and amount of (possibly recurring) donations, and enhancing the NPOs' brand advocacy.%

\section*{Acknowledgements}

This research was conducted by the ARC Centre of Excellence for Automated Decision-Making and Society (project number CE200100005) in collaboration with the Australian Red Cross, and funded partially by the Australian Government through the Australian Research Council.%

\printbibliography

\end{document}